\documentclass{aastex63}
\submitjournal{AJ}
\pdfoutput=1
\begin{document}
\title{The Dynamics of the Wide-Angle Tailed (WAT) galaxy cluster Abell 562}

\author{P.L. G{\'o}mez}
\affiliation{ W. M. Keck Observatory, 65-1120 Mamalahoa Highway, Kamuela, HI 96743, USA }
\author{D. Calder\'on}
\affiliation{Institute of Theoretical Physics, Faculty of Mathematics and Physics, Charles University, 180 00 Prague, Czech Republic}

\begin{abstract}

We present the first in depth dynamical analysis of the archetypal Wide-Angle Tailed (WAT) cluster Abell 562. We have combined Gemini observations with archival data from the literature to form a sample of 76 cluster members and derived a mean redshift of $0.1088 \pm 0.0004$ and a velocity dispersion of $919 \pm 116$ km
s$^{-1}$. This relatively large velocity dispersion suggests either a very massive cluster ($M_{dyn} > 6.9 \times$ 10$^{14}$ M$_\sun$) and$/$or a merger system.  The merger model is supported by a non-Gaussian galaxy velocity distribution, an elongated spatial distribution of likely cluster members, and an elongated X-ray emitting gas. This scenario would generate the bulk flow motion of the intra-cluster medium that can exert enough ram pressure to bend the radio jets. Thus, our observations support the model in which a recent off-axis merger event produced the cluster wide conditions needed to shape the WAT in Abell 562.

\end{abstract}

\section{Introduction}
\label{sec:intro}

Cluster of galaxies grow and evolve by continuous accretion of mass through merger events. These mergers produce physical effects on the intra-cluster medium (ICM) such as bulk flows of gas, shocks, and the formation of substructure in the gas density and temperature. This ``stormy weather'' \citep{Science..280..400} can be responsible for the morphology and perhaps the formation of the so-called Wide-Angle Tailed (WAT) radio sources. 

WAT radio sources \textbf{were first classified as such by \citet{1976ApJ...205L...1O} based on their particular bent morphology}. They have radio powers are at the limit between the very powerful FR IIs and the weaker FR Is. The size of the radio lobes extends beyond the limits of the galaxy host, and thus, makes them useful probes of the intra-cluster medium (ICM). In addition, the hosts for the WATs are central dominant galaxies (D or cD galaxies, \citet{1976ApJ...203L.107R}) which are expected to either be at rest or have small peculiar motions in the cluster. Otherwise these galaxies could be destroyed by tidal forces. 

Several models have been proposed for explaining their peculiar radio morphology. \textbf{Early models proposed that radio bent sources are produced by individual plasma blobs that are ejected by the AGN and adopt the observed geometry due to buoyancy or interactions with the ICM \citep{1973A&A....26..423J, 1978ApJ...223L...9F}.  However, \citet{1979AJ.....84.1683B} found that the physical properties of these blobs were not consistent with the observations of the WAT 1159+583. A slight variation of this model was proposed by \citet{1984MNRAS.208..323L} who postulated that the AGN jets left trails that in their wake adopted the observed bent morphology. Other models assumed that the observed morphology is produced not by blobs but by actual AGN jets that are bent by buoyancy and pressure imbalances \citep{1997A&A...327...37S}, instabilities at the ISM/ICM boundary \citep{1986MNRAS.220..351P}, electrodynamic effects \citep{1984ApJ...278...37E} or ram pressure \citep{1979Natur.279..770B, 1985A&A...147..321F, 1993ApJ...408..428O}. \citet{1996MNRAS.283..673S} proposed a combination of ram pressure and buoyancy to explain the complex morphology of the WAT in 4C 34.16.}

\textbf{In order for the ram pressure model to work, there must be sufficient relative velocity between the AGN and the ICM.} The estimated relative velocity that is needed to produce the bending of the jets depends on fundamental jet properties such as velocity and density. For instance, \citet{1997AJ....114.1711G} estimated that relative velocity between 400-2500 km s$^{-1}$ would be needed for slow jets ($v_j \sim 0.0075c$). \citet{2006MNRAS.368..609J} expanded this analysis to faster jets ($v_j \sim 0.6c$) and concluded that lighter jets might need even lower galaxy velocities. 

The hosts for the WATs are central dominant galaxies which are expected to be at rest or have very small peculiar motions in the cluster. So if the host galaxy does not have a significant peculiar velocity what could be the source of the large relative velocity between the jets and the ICM needed to bend the WATs? The most accepted scenario is that the ram pressure is produced by cluster mergers. There is evidence suggesting that most WAT sources are located in clusters that show the effects of recent mergers such as substructure in the X-ray emitting gas \citep{1997ApJ...474..580G}, the galaxy distribution, or the presence of non-Gaussian velocity distributions  \citep{2000AJ....120.2269P}. Moreover, numerical simulation of cluster mergers predict that large bulk flow motions develop during a merger. They are produced just after the core crossing and can last for as long as 1 Gyr \citep{1996ApJ...473..651R, 1995ApJ...445...80L}. These flows develop along the merger direction and are parallel to the elongations of the X-ray emitting gas that are produced during the merger. \textbf{Even though this model seems to explain most WATs, there are a few counterexamples (e.g., the WAT in 3C 130 as described by \citet{1998MNRAS.298..569H}) that merit further work.}

Abell 562 \citep{1989ApJS...70....1A} is a nearby cluster that hosts a prototypical WAT radio source \citep{1997ApJ...474..580G, 2011ApJ...743..199D}. This radio source is also known as 3C169 \citep{1961ApJ...134..970V} or 0647+693 \citep{1985AJ.....90..954O} and exhibits the typical \textbf{bent} shape of its twin jets. \citet{2006MNRAS.368..609J} calculated that a galaxy velocity of 870 km s$^{-1}$ would be needed to bend this source based on the radio properties of the jet and the X-ray properties of the surrounding gas. They argued that such a velocity is probably too large for a centrally located galaxy and can only be produced by a cluster merger. In fact, \citet{2002AJ....124.1918M} reported a recessional velocity for this galaxy of 32,448 km s$^{-1}$ and a cluster recessional velocity of 32,758 km s$^{-1}$ based on the observations of 6 other cluster members. These measurements reveal a small peculiar velocity of 310 km s$^{-1}$ for this WAT and support the merger hypothesis.

\cite{2011ApJ...743..199D} have analyzed Very Large Array (VLA) and Chandra observations that further support the merger model. Their Chandra analysis confirmed an elongated X-ray emission along the line that bisects the jet bending which was first detected in ROSAT data by \cite{1997ApJ...474..580G}. Moreover they reported that the X-ray temperature spatial distribution shows signatures of a recent cluster merger. Thus, they found regions of hot gas probably caused by shock heating. Finally, they found that the excess X-ray emitting gas located between the WAT lobes has a high abundance and proposed that it might be a remanent cool core disrupted by a merger.  

In this paper, we combine recent recessional measurements of 75 cluster members with archival data of one cluster member in order to confirm the small peculiar motion of the WAT. Furthermore, we use these data for conducting statistical analysis in order to test the merger hypothesis and model the type of merger experienced by this cluster. In this way, we can further constraint the relative velocity and the jet properties for this prototypical WAT radio source. We report  our optical observations in section \ref{section:optical} and their data analysis in section \ref{section:opt-analysis}. Our interpretations are discussed in Section~\ref{section:model}. Finally, we summarize our findings in Section~\ref{section:conclusions}.  Note that we use $H_0 = 70$ km s$^{-1}$ Mpc$^{-1}$, $\Omega_\Lambda = 0.714$ and a flat universe throughout the paper, so that 1\arcmin $\simeq$ 119.8 kpc. All errors are quoted to the 1$\sigma$ level.

\section{Observations}
\label{section:optical}

In this section, we report the optical and X-ray observations of Abell 562. We
start by describing the pre-imaging data used to design the multi-object spectroscopic (MOS) masks. Next, we describe the GMOS-N MOS spectroscopic observations and their reduction. Finally, we describe our analysis of archival Chandra X-ray observations that will be used to compare the distribution of the galaxies with the distribution of the X-ray emitting gas. 

\subsection {Optical Imaging}

Imaging data in the $g'$ (120 seconds total) and $r'$ bands (120 seconds total) were obtained with the GMOS instrument mounted at the Gemini North telescope. We used four different pointings aimed at Abell 562 in order to map out the galaxies located within $\sim$ 5\arcmin $\:$ of the cluster center. The observations
were carried out in queue mode (program GN-2008B-Q-97 and GN-2009B-Q-46) with an overall seeing of $\sim$ 1.0\arcsec $\:$ and thin cirrus on the sky. The images were bias subtracted and flat fielded with sky twilight flats.  Next, the GMOS images were mosaiced and coadded into final $g'$ and $r'$ band images. We identified and measured the properties of the galaxies using the SExtractor program
\citep{1996A&AS..117..393B}. We adopted the MAG-BEST estimate as the
measurement of the galaxy magnitude. Even though the observations were
not obtained under photometric conditions, we converted the observed
counts into approximate magnitudes using the zero-point filter values
listed on the Gemini public webpages. We are aware that the absolute
magnitudes could be off by up to $\sim$ 0.3 magnitudes. We
combined the SExtractor output from the $g'$ and $r'$ images to
construct a color-magnitude diagram. The color term should not be severely affected by the uncertainty in the absolute magnitude of the galaxies. Figure \ref{fig:colmag} shows the
color-magnitude diagram and Figure \ref{fig:galdist} shows the
distribution on the sky of 821 objects derived from the SExtractor
output. All of these objects are likely to be galaxies as they had the
CLASS-STAR $<$ 0.9. We selected a sub-sample of 481 of these 821 as
candidate cluster members based on their position on the
color-magnitude diagram ({\it i.e.} $0.3  <  g'-r' < 1.7$, and $m_{r'} < 18.6$). This color cut is wide enough to include all the red sequence galaxies and other likely cluster members.

Note that we also observed the core of the cluster with the DEIMOS instrument at W. M. Keck Observatory in the $V$ (900 seconds total time), $R$ (900 seconds total time) , and $I$ bands (540 seconds total time). These observations were obtained in March 30, 2017 under clear conditions and 0.8 arcsec seeing. These images will be used to look for multiple nuclei in the WAT host.

\subsection{Optical Spectroscopy}
\label{sec:opt-spec}

We observed 131 unique targets in Abell $562$  with the Gemini North Multi-Object Spectrograph (GMOS-N) in MOS mode as part of the queue program GN-2010B-Q-90 in November 2010. Each mask had a field of view of $\sim 5\arcmin \times 5 \arcmin$ that allowed us to fit at least 30 candidate galaxies. In order to observe at least 100 galaxies, we designed 4 GMOS masks covering a region $\sim 10 \arcmin \times 10 \arcmin$. The sky coverage of the 4 masks can be seen in figure \ref{fig:galdist}. In our spectroscopic setup, we used the B600+G5307 grating that has an average spectral resolution of $\sim$180 km s$^{-1}$ ($R\sim1700$) or $\sim0.5\AA$ per pixel  centered between 520-530 nm. These wavelengths correspond to the spectral location of the redshifted Ca[II] absorption lines of early type galaxies which constitute the bulk of our cluster candidates. The observations consisted on two 1800 sec exposures spectrally shifted to minimize the effects of cosmic rays, bad pixels, and chip gaps.  

To reduce the spectroscopic data we used the Gemini $IRAF$ package and followed the standard reduction recipe. First, we bias corrected and flat fielded the data, removed cosmic rays and mosaiced the science and arc frames. We wavelength calibrated the arc frames row by row and visually inspected the corrected arc images before applying the calibration to the science data. Then, we defined the apertures for each galaxy and extracted and sky-subtracted 1D spectra from the 2D spectra. Finally, the two 1800 seconds spectra were co-added by shifting them to a common wavelength frame. Typical error of the wavelength calibration was about 0.5\AA. 

To measure the radial velocities of the observed galaxies, we used the tasks $fxcor$ from $RV$ and $emsao$ from $RVSAO$. $Fxcor$ cross-correlates the observed spectra with high signal-to-noise spectra templates for different galaxy types obtained from the SDSS database. On the other hand, $emsao$ finds emission lines, computes redshifts for each identified line, and combines them into a single radial velocity. The typical error for radial velocities (including the error due to wavelength calibration and the trace) was $\lesssim$ 100 km $s^{-1}$. Note that we observed 11 objects twice. The mean difference between the measurements of these 11 repetitions is 75 km s$^{-1}$ and confirms the velocity error quoted above.

\subsection{Chandra X-ray Observations}
\label{sec:xray}

Abell 562 was observed by the Chandra ACIS-S instrument in 2006. We retrieved the data (obsid 6936) from the Chandra Data Archive (CDA) and analyzed them with CIAO version 4.10. We filtered the data by analyzing the light curves on the S3 chip. Thus, we ended with an actual exposure time of 51,392 sec. Next, we masked out point sources and created exposure maps in order to map the extended cluster X-ray emission in the 0.2-10keV spectral region. As expected, the final map (see figure \ref{fig:xrayimage}) shows the same properties initially reported by \cite{2011ApJ...743..199D}.

\section{Analysis of the Optical Data}
\label{section:opt-analysis}

\subsection{Completeness of the Spectroscopic Sample}
\label{sec:completeness}


After reducing and processing the spectroscopic data, we were able to measure the velocities for 102 objects 
 (Table \ref{tab:galaxies}).  Although our sampling is not complete, we believe that it is representative of the galaxy population in Abell 562. For example, of the galaxies targeted for
spectroscopy, we have been able to observe over 50$\%$ of those brighter than $r'= 17.$ (and roughly 30$\%$ at $r'=18$).  We have also compared the radial distribution of the observed galaxies with the distribution of all the 481 galaxies selected as candidate cluster members (see \S~\ref{sec:opt-spec}).  We computed the ratio of the number of observed galaxies to the number of candidate members in 3 radial bins, each 2.0$\arcmin$ ($\sim$ 225 kpc) wide, and centered on the brightest galaxy in the cluster. We obtained values of 0.45, 0.27 and 0.29 for the ratios in the three bins. This distribution shows that we selected relatively
more galaxies toward the center of the cluster. This is expected because the four masks slightly overlap towards the center of the cluster (see figure \ref{fig:galdist} and figure \ref{fig:xrayimage}).

\subsection{Mean Redshift and Velocity Dispersion}

The 102 galaxies with redshifts in Table~\ref{tab:galaxies} include cluster members, foreground, and background galaxies. In order to increase the number of galaxy velocities available for our dynamical study we combine this sample with data for 11 galaxies obtained from the literature \citep{2002AJ....124.1918M}. We find that our measurements and the data from \citet{2002AJ....124.1918M} have two galaxies in common. The mean difference between these measurements is 130 km s $^{-1}$ which is comparable with the average error in the measurements and do not reveal a systematic difference. Since we plan to test for substructure by combining the the spatial and velocity data, we decided to apply a velocity and a spatial filter. First, we only consider galaxies located within a radius of $7 \arcmin$.   Next, we used an iterative 3 $\sigma$ clipping criterion
\citep{1977ApJ...214..347Y} to identify the cluster members. Most of the outliers were rejected after the first iteration. Thus, we
were left with 76 likely cluster members. The velocities for 75 of these galaxies were derived from our own observations (labeled as such in Table \ref{tab:galaxies}); the velocity for one galaxy was  obtained from the literature (with velocity of 33,043 km $s^{-1}$ for a galaxy located at RA $=$ 06 50 38.27 and DEC$=$ 69 24 19.9 from \citet{2002AJ....124.1918M}. The density of cluster candidates and spectroscopically confirmed members, versus position in the X-ray image, is
shown in Figures ~\ref{fig:galdensityc} and \ref{fig:galdensityv} respectively. As we are dealing with small
number statistics, we used robust bi-weight estimators for the mean and velocity dispersion in order to characterize the distribution of
redshifts \citep{1980A&A....82..322D}. We used the astropy \citep{2013A&A...558A..33A,2018AJ....156..123A} implementation of the ROSTAT program \citep{1990AJ....100...32B} biweight estimators. We measured a bi-weight estimator for the location of the cluster of $32,643 \pm 120$ km
s$^{-1}$ (1 $\sigma$ errors) with a velocity
dispersion (cosmologically corrected) of $ \sigma_{vel}= 919 \pm 116$ km
s$^{-1}$ (1 $\sigma$ errors determined with a bootstrap method). We note that the measured cluster redshift of 0.1088 is in agreement (within the errors) with
the 0.1092 value quoted by \citet{2002AJ....124.1918M}. Its location on the sky and its redshift are shown
in Figures~\ref{fig:galdist} and \ref{fig:velhist} respectively.

We note that due to the fact that the GMOS field only samples the inner 7$\arcmin$ (or 839 kpc) of the cluster the velocity dispersion could be biased and
overestimate the actual value by as much as
 10$\%$ (see \citep{2013ApJ...772...47S}). Therefore, in what follows, we will use a conservative correction factor of
$\sim 10\%$ to the velocity dispersion so that the effective velocity dispersion $ \sigma_{vel,eff}=$ 827 km
s$^{-1}$.

\subsection{WAT Peculiar Motion}
\label{subsubsec:watpec}

We measure a negligible peculiar motion of the WAT with respect to the cluster location. The recessional velocity of the WAT is $32,750 \pm 60$ km
s$^{-1}$ and the cluster location has a value of $32,643\pm 120$ km s$^{-1}$.  Thus, the peculiar velocity is $107 \pm 134$ km s$^{-1}$ (corrected to the cluster rest frame). Thus, the WAT galaxy is at rest with respect to the cluster. This is represented on Figure~\ref{fig:velhist} that shows the velocity of the WAT and the cluster location with their $99.7\%$ error bar.

\subsection{Cluster Mass}
\label{subsubsec:opticalmass}


We have computed the mass of the cluster by using the $M_{dyn}$ relation from \citet{2013ApJ...772...47S}
:

\begin{equation}
M_{dyn}= [\frac{\sigma_v}{A \times h_{70}(z)^C}]^B \times 10^{15} M_\sun
\end{equation}

where $M_{dyn}$ is the dynamical mass within a virial radius and the constants used are: $A = 939$, the exponent $B = 2.91$, and the exponent $C = 0.33$ according to \citet{2013ApJ...772...47S}. In our case, we use the effective velocity dispersion of $ \sigma_{vel,eff}= 898 \pm 220$km
s$^{-1}$ and calculate a value of $M_{dyn} = 6.9 \pm 2.8 \times$ 10$^{14}$ M$_\sun$.


\subsection{Dynamical State}

\subsubsection{Substructure}
\label{subsubsec:structure}

In order to determine the dynamical state of the cluster, we examined the velocity and spatial distribution of the galaxies using 1-d, 2-d and 3-d statistical tests. The 1-d statistical
tests look for non-Gaussianity and $/$or substructure in the velocity distribution. The 2-d tests look for asymmetries and substructure in the spatial distribution of the galaxies. Finally, the 3-d tests, which combine the velocity and position information, look for other merger signatures.

We run a modified version \citep{1996ApJS..104....1P} of the ROSTAT \citep{1990AJ....100...32B} battery of tests on
the velocity distribution (Figure~\ref{fig:velhist}) to quantify any non-Gaussianity. Kurtosis is detected by the $W^2-stat$ with a probability of Gaussianity of $4\%$, $A^2-stat$ with a probability of $5\%$, and $U^2-stat$ with a probability of $3\%$. The velocity distribution appears asymmetric. Thus, we measure an asymmetry index $AI$ \citep{1993AJ....105.1596B}  -0.7 that has a probability of $8\%$ of being Gaussian and a Tail Index $TI$ \citep{1993AJ....105.1596B} value of 1.4 that has only a probability of 1$\%$ of being Gaussian. Therefore, we found evidence for non-normality in the velocity distribution of the galaxies.

Next we examined the galaxy distribution for evidence of substructure. Figure~\ref{fig:galdensityv} shows the spatial distribution of the spectroscopically confirmed galaxies. Their distribution shows a central peak and two elongations. The central peak coincides with the peak of the X-ray emission. One elongation is towards the NE and the other is towards the SSW. In order to further explore this peculiar distribution, we have analyzed the spatial distribution of the most likely cluster members as selected from the color-magnitude diagram. Thus, we have selected 316 galaxies located within $\pm 0.3$ magnitudes of the red sequence ($i.e.$, color-selected sample or CS sample). Their distribution is shown in figure \ref{fig:galdensityc}. It is very similar to the distribution of the spectroscopically confirmed members  in figure \ref{fig:galdensityv}. It shows a central clump elongated towards the NE (parallel to the elongation found in figure \ref{fig:galdensityv}.  This clump is slightly offset from the X-ray peak. It also shows an elongation towards the S which is similar to the one seen in figure \ref{fig:galdensityv}.

We assess the significance of the substructure by running a
series of 4 tests on the CS sample: The angular separation \citep{1988ApJ...327....1W} test (i.e, AST); the so called $\beta$ test
\citep{1988ApJ...327....1W}; the Fourier elongation test \citep{1996ApJS..104....1P}; and the Lee Statistic
\citep{1979ApJ...229..424L, 1987ApJ...317..653F}. Of these, only the Lee statistics shows a marginal probability of substructure as seen in table \ref{tab:tab2d}. This test looks for a significant split into two subsamples and confirms the bi-modality hinted by figure \ref{fig:galdensityc}. We did not run the tests on the sample of confirmed members for two reasons. First, the color selected sample has at least three times more galaxies. Second, the spectroscopic sample contains galaxies preferentially located towards the cluster center due to the central overlap of the four masks as shown in figure \ref{fig:galdist} in section \ref{sec:completeness}.

Finally,  we looked for substructure
using three-dimensional tests that  combine the spatial and
kinematical positions of the galaxies. Please note that our spatial sampling of the galaxies is not uniform and this could
affect the sensitivity of these tests. We ran 3 different
tests on the data: the Lee 3-d test \citep{1996ApJS..104....1P}, the
Dressler-Schectman test \citep{1988AJ.....95..985D}, the $\epsilon$ test \citep{1993PASP..105.1495B}, and the $\alpha$
test \citep{1990ApJ...350...36W}. These tests did not find any significant evidence of substructure as reported in table \ref{tab:tab3d}.

\subsubsection{Clustering}

It is possible that the substructure tests run before (\ref{subsubsec:structure}) are insensitive to some types of mergers and clustering in the data. For instance, \citet{1996ApJS..104....1P}  reported that these tests were very effective at detecting substructure and that was the reason we initially used them to analyze our data. However, we note that \citet{1996ApJS..104....1P} evaluated these tests using only a specific type of merger: simulated \textit{head-on} mergers. Could these substructure tests be less effective in the case of non-head on mergers? In order to be more thorough, we turned to use exploratory data analysis (EDA) as an alternative method to look for any previously undetected clustering in the data. Specifically, we used the astropy implementation of the Gaussian mixture models (GMM) as an additional method for objectively identifying galaxy densities especially for the 3D case that combines spatial and velocity information. This method was first used by \citep{1994AJ....108.2348A} to detect bimodality on astronomical data sets. Since then the method has been used to look for substructure in individual clusters (e.g. {\citep{1998AJ....115....6B, 2017A&A...602A..20B, 2020MNRAS.492.2405B}). 

If this cluster is undergoing a merger, hierarchical evolution suggests a bimodal cluster merger. Therefore, we  concentrate  on the GMM model that splits the data into two groups: group A with 56 members and group B with 20 members.  Table \ref{tab:tabemm} shows the properties of these two groups. Figure \ref{fig:2demm} shows overlays of the group A and B galaxy densities over the smoothed X-ray data. Interestingly, group A is compact and elongated and it is aligned with the main X-ray elongation. Group B on the other hand shows a peak to the South of the main group and is less concentrated than group A. Figure \ref{fig:3dall} shows a 3D scatter plot of the galaxies in spatial and velocity space. We also project the histograms of these groups. The galaxies in group A are more concentrated spatially and in velocity space  than galaxies in group B. Moreover, group A is located at a slightly higher redshift than group B. \textbf{Finally, the GMM allocates the WAT to galaxy group A with a probability greater than 0.99. In this case, the relative peculiar velocity of the WAT with respect to group A would also be negligible (89 $\pm 135$ km s$^{-1}$)}

It is possible that these two groups trace the pre-merger components. Based on the velocity dispersion of them we can derive a mass ratio of 4:1 (based on their velocity dispersions). Another way is that the groups do not trace the pre-merger component. Thus, group A points to the highest galaxy density clump probably composed on the merging cores of the two merging clusters. Whereas group B  is made up of the galaxies that have not mixed yet and are still trailing the cores of the pre-merger systems. 

It is also possible that the GMM method can split the cluster into more than two components. Therefore, we used both the Akaike and Bayesian information criteria (AIC and BIC)  to determine if splitting the data into more than two components is statistically significant.  These criteria are based on combining the maximum likelihood computed for each model  and correcting it by a function of the number of components. This is to prevent the condition at which the maximum likelihood is obtained when the number of components is equal to the number of elements. The best model would minimize these criteria. Figure \ref{fig:ABstat} shows the BIC/AIC criteria as a function of number of components for our 3D data. Even though these two criteria do not single out a bimodal model as the most statistically significant, they do prefer models with few components. 

\subsubsection{Dynamical State}

In summary, the strongest optical evidence for a
merging event comes from the skewness of the velocity distribution and from the galaxy density
distribution. Figures~\ref{fig:galdensityv} and \ref{fig:galdensityc} show
an elongation and bi-modality consistent with that seen in the X-ray image. This substructure is confirmed by
the Lee statistical test when applied to the color selected sample.  It is
worth mentioning that there is no single test that can determine the
dynamical state of every cluster \citep{1996ApJS..104....1P}. Therefore, it is not surprising that some tests yield inconclusive results. After all, the sensitivity of a given
test to detect substructure (or other perturbations such as asymmetry)
depends on both the quality and completeness of the data available and
on the properties of the merging system, such as mass ratio, merger
epoch, and viewing angle ({\it e.g.} \citep{1996ApJS..104....1P, 2010MNRAS.408.1818W}). Interestingly, the GMM method finds clustering in the data and the GMM model with two components finds alignments between galaxy and gas clumps. Therefore, a more insightful approach would be to compare the observables with numerical simulations. This approach has already been applied with success to other clusters ({\it e.g.} \citep{2007MNRAS.380..911S}). We describe an initial qualitative attempt at this type analysis in the next section.

\section {Discussion}
\label{section:model}

Abell 562 shows signatures of a cluster merger observed after core crossing. The spatial distribution and the thermal properties of the X-ray emitting gas are consistent with a post merger cluster.  In their Chandra analysis, \citep{2011ApJ...743..199D} reported an elongation of the X-ray emitting gas aligned with the line that bisects the jets of the WAT. Moreover, the distribution of gas in the core of the cluster deviates from an elliptical 2D $\beta$-model due to the presence of  a clump of high metal abundance gas located in between the radio lobes.  Interestingly, they proposed that the high metal abundance clump of gas might be the remanent of one of the pre-merger cluster cores. The substructure and elongation of the X-ray emitting gas and the alignment between the X-ray and the WAT jets are the typical signatures of a recent cluster merger.  Note that a relative gas velocity of 1000 km $s^{-1}$ is what is needed to bend the jets \citep{2011ApJ...743..199D}. This type of bulk flow gas motion is produced during a merger and can only lasts for a couple of Gyrs after core crossing \citep{1996ApJ...473..651R}. \textbf{Moreover, we have measured a small relative velocity between the WAT host and the rest of the galaxies. Therefore, its contribution to the overall relative velocity between the WAT host and the ICM is negligible.}

Our substructure analysis of the galaxy data is also consistent with a merger model. During a merger and depending of the merger parameters (i.e., epoch, viewing angle, and mass ratios)  the velocity of the galaxies could show skewness and the spatial distribution could be elongated \citep{1996ApJS..104....1P} . In the case of Abell 562, we have shown that the elongation of the spatial distribution of the galaxies is aligned with the elongation of the X-ray emitting gas. Depending on the merger parameters it might be possible to identify some of the pre-merger parameters. For instance, during core crossing the spatial distribution of the galaxies would be mixed whereas the galaxies would be segregated in velocity space. This is due to an ever decreasing center of mass separation and an increasing relative velocity before core-crossing (maximum infall velocity) and even after core-crossing (as the pre-merger cluster galaxies oscillate towards a relaxed state).  This will hold for most viewing angles and for a time interval of $\sim$2Gyr before and after core crossing depending on the mass ratio and impact parameter of the merger. On the other hand, observing a pre-merger system would show the opposite signatures because the galaxies would still be clustered around their center of mass but their velocity distributions would overlap as they have the same redshift. 

Based on the evidence for a merger in Abell 562 we decided to compare the data with simple N-body simulations of cluster mergers. We refer the reader to \cite{ 2002ApJ...569..122G} for details about the simulations and scaling of the simulations with real data. We concentrated on head-on and non head-on mergers (with impact parameters from 250 kpc to 750 kpc) and mass ratios from 1:2 to 1:4. In order to compare the simulations to the observed velocity distribution of the galaxies it was necessary to apply a numerical scaling to the velocities of the N-body particles, since the model calculations were done in scale-free coordinates. We used the Kolmorov-Smirnoff (KS) test to estimate the probability that the observed  galaxy velocity distribution and the velocity distribution of the N-body particles (sampled during a merger) are drawn from the same parent population.

Some of the results are  shown in Figure \ref{fig:accsim}. In this figure the vertical axis shows the initial mass of the main merging subcluster while the horizontal axis shows the time since closest core approach for a 1:4 mass ratio mergers and 400kpc impact parameter. The different symbols
indicate the different viewing directions assumed. Only models with a (2-sided) KS probability for rejection of 70$\%$  or more are plotted. Overall 8700 possible models as a function of epoch (from $\sim$6 Gyr to 5 Gyr with a typical timestep of $\sim$0.3 Gyr) and initial main subcluster mass (from $0.7 \times 10^{15} M_\sun$to $1.5 \times 10^{15} M_\sun$ with a typical mass spacing of $\sim 4 \times 10^{13} M_\sun$) were sampled for each viewing angle.  Figure  \ref{fig:plot_histsim} shows the velocity distribution envelope of the models with KS probability for rejection of 70$\%$  or more and shows how these "accepted" models compare with the observed velocity distribution. These results demonstrate that the observed velocity distribution in A562
is consistent with the velocity distribution expected from the major non head-on merger of two subclusters (mass ratio 1:4), close to or after the time of core-crossing, with the exact epoch depending on the mass ratio and viewing geometry. In addition, figure \ref{fig:contsim} is an example of the projected dark matter surface model for one of these "accepted" models that is consistent with the spatial bimodality revealed by the GMM analysis. This is not an attempt to find the best model but only to explore if a non head-on merger could provide a reasonable model for the observations. \textbf{Finally, note that the WAT host could be at rest relative to the bulk of the cluster during this merger because the merger is non head-on and it is very asymmetric (4:1 mass). The main assumption here is that the WAT galaxy probably originated in the most massive merging cluster and close to the bottom of its potential well. This assumption is supported by the GMM which allocated the WAT host to the most massive galaxy group A. Unfortunately we can not confirm this with our N-body simulations because they lack the mass and spatial resolution to identify individual galaxies.} 

At this time, our simulations do not include gas. But we have looked at the literature to explore the properties of the cluster gas during a similar merger. We found a previous detailed hydrodynamical numerical simulation of a non-head on merger for Abell 754 \citep{1998ApJ...493...62R}. Figure 
1 of that paper shows an X-ray emission with two elongations. The main X-ray emission is elongated in the  NNE to SSW direction whereas there is another fainter large scale elongation in the SE to NW direction. The spatial distribution of the galaxies is also bimodal. One clump is located on top of the main X-ray emission (labeled as SE) whereas the second clump is located some 900 kpc away towards the edge of the large SE to NW X-ray emission (labeled NW). The most interesting characteristic is the fact that the small scale X-ray gas elongation is not parallel to SE to NW direction that connects the two galaxy clumps. The presence of two groups of galaxies aligned in a direction that is not parallel to the main X-ray elongation is also present in Abell 562. 

The hydrodynamical simulation reported by \citet{1998ApJ...493...62R} (also see \citet{1996ApJ...466L..79H}) models Abell 754 as a post-merger non head-on merger (impact parameter of 120 kpc) of two clusters with mass ratio of 2.5:1. Moreover, they propose that the merger occurred close to the plain of the sky and is being observed some $\sim$ 0.5 Gyr after core crossing. Thus, a similar non head-on merger would be consistent with most of the acceptable models for Abell 562 (see figure \ref{fig:accsim}). In addition, these hydro/N-body simulations provide additional information about the velocity properties of the cluster gas at different epochs. For instance, Figure 6 shows the distribution of the gas velocity some 2.75 Gyrs after the moment of closest approach. This is later than most of the models that we have proposed  for Abell 562 (peak at 2 Gyrs after the epoch of closest approach). However, we can derive some insightful  information about the kinematical properties of the X-ray gas.  For instance at  0.3 Gyrs the bulk flow velocity peaks at over 1800 $km s^{-1}$ and at 2.75 Gyrs it peaks at over 950 $km s^{-1}$. As described by \cite{2011ApJ...743..199D} these are the bulk flow magnitudes needed to bend the radio jets and form the WAT.

\subsection{WATs as Wind Socks}

\textbf{In the previous sections we have described a merger model that can explain the main bending observed in the tails of the WAT radio source. However, this source shows other morphological features whose origin is still unexplained. For instance, }the jets in 0647+693 show several minor bends (i.e., N1, N2, and N3 in Figure 11 of \cite{2011ApJ...743..199D})  and changes in direction. As shown by \cite{1998ApJ...493...62R}  a non-head on merger induces significant angular momentum to the cluster gas that manifests itself as rotations and eddies. Therefore, we are wondering if these bends and twists could be produced by the interaction between the jets and the different type of gas flows stirred by the merger. In order to make a qualitative assessment, in figure \ref{fig:plot_merged} we have overlaid the radio image of the WAT on the hydro/N-body velocity fields produced by \citet{1998ApJ...493...62R} for the A754 merger at 0.5 Gyr after the epoch of closest approach. It is tantalizing to see that the N1, N2, and N3 bends can be naturally caused by ram pressure produced by the rotating and turbulent gas.  This is only a qualitative comparison that can motivate further studies of the pressure balances along the jets and comparison with the velocity fields produced by Hydrodynamical cluster merger simulations. Could this suggest that WAT radio sources with severely distorted tails (i.e. A562) are only present in non-head mergers whereas bent radio sources with symmetric tails (e.g., Abell 2634 as described in \cite{1993ApJ...416...36P}) are mostly present in head-on mergers? We plan to further explore the potential of the radio morphology as a merger diagnostics on larger samples of WAT clusters with high spatial resolution VLA maps. Finally, we look forward to map out cluster gas velocities with the upcoming XRISM mission and compare them with the WAT geometry.

\section{Conclusions}
\label{section:conclusions}

We measured new redshifts for 102 galaxies in the vicinity of Abell 562 with the GMOS-N instrument mounted on the Gemini North telescope. After combining with data in the literature, we obtained a sample of 76 likely cluster member galaxies. We concentrate our study of the cluster's kinematics on these galaxies.

We calculated a new robust redshift for this cluster of 0.1088 and a velocity dispersion of $919\pm116 ~km s^{-1}$. Moreover, we found a negligible peculiar velocity for the WAT galaxy which is typical of these radio sources.

We find evidence for kinematic structure and non-Gaussianity in the velocity data of these galaxies. The velocity distribution is asymmetric and shows kurtosis. Next, we find some evidence of substructure as quantified by the Lee statistics on the sample of color selected galaxies. Even if our 3D statistical tests do not find evidence for substructure, we find interesting results when we use the Gaussian mixture modeling on the data. The GMM technique used as a density estimator splits the data in two statistically significant groups. 

Motivated by these results and the analysis of Chandra X-ray data from the literature, we compared our spatial and velocity data with simple N-body simulations of non head-on cluster mergers. We find that the velocity distributions produced by the merger of two sub clusters with mass ratio of about of 1:4 near the time of core-crossing and occurring close to the plane of the sky are consistent with the observed velocity distribution. In addition, near the epoch of core-crossing the 1:4 merger produces a qualitatively similar X-ray spatial distribution as the one seen in Abell 562.

In summary, our kinematical study supports the view that Abell 562 is a non head-on merging cluster and that the WAT galaxy does not have a significant peculiar velocity. These findings support the hypothesis that the WAT bending is produced by the interaction of the radio jets with the bulk flow motions of gas produced by a merger. Further insights into the nature of this WAT and into the process of cluster merging should be forthcoming with the data expected from the new generation of X-ray missions. 

\acknowledgements
{We would like to thank the anonymous referee for the comments made to the draft. They helped us to clarify and better focus the paper. We would also like to thank E. M. Douglass for providing us with the 1.4GHz radio data. This research was possible with financial support from: This project was conducted in the framework of the CTIO PIA Program.
This is a CTIO funded summer student program ran in parallel with the CTIO REU Program, which was supported by the National Science Foundation under grant AST-1062976. This research has made use of data obtained from the Chandra Data Archive and software provided by the Chandra X-ray Center (CXC) in the application packages CIAO.This work is based on observations obtained from: {\it a)} the {\bf
  Gemini Observatory}, which is operated by the Association of
Universities for Research in Astronomy, Inc., under a cooperative
agreement with the NSF on behalf of the Gemini partnership: the
National Science Foundation (United States), the National Research
Council (Canada), CONICYT (Chile), the Australian Research Council
(Australia), Ministerio da Ciencia e Tecnologia (Brazil) and SECYT
(Argentina); {\it b)} the {\bf Chandra X-ray Observatory}, Chandra is
operated by the Smithsonian Astrophysical Observatory for and on
behalf of the National Aeronautics Space Administration under contract
NAS8-03060. {\it c)} the {\bf W. M. Keck Observatory}, which is operated as a scientific partnership among the California Institute of Technology, the University of California and the National Aeronautics and Space Administration. The Observatory was made possible by the generous financial support of the W. M. Keck Foundation. The authors wish to recognize and acknowledge the very significant cultural role and reverence that the summit of Maunakea has always had within the indigenous Hawaiian community.  We are most fortunate to have the opportunity to conduct observations from this mountain. }

\facilities{Chandra(ACIS)}{Gemini(GMOS)}{Keck(DEIMOS)}

\clearpage

\begin{deluxetable}{c c c c c}
\tablewidth{0pt}
\tablecaption{\label{tab:specsetup} GMOS-South Spectroscopic Setup}
\tablehead{
  \colhead{mask}& \colhead{grating ruling density}& \colhead{central
    wavelength}& \colhead{full range}& \colhead{blaze
    wavelength}\\
  \colhead{}& \colhead{lines mm$^{-1}$} & \colhead{\AA}&
  \colhead{\AA}& \colhead{\AA}}
\startdata
mask1 & 600 & 5300, 5500 & 2760 & 4610\\
mask2 & 600 & 5300, 5500 & 2760 & 4610\\
mask3 & 600 & 4600, 4800 & 2760 & 4610\\
mask4 & 400 & 7800, 8000 & 4160 & 7640\\
\enddata
\end{deluxetable}
\startlongtable
\begin{deluxetable}{c c c c c c}
\tabletypesize{\footnotesize}
\tablewidth{0pt}
\tablecaption{\label{tab:galaxies}  Spectrocopic Targets}
\tablehead{
\colhead{id}& \colhead{RA(2000)} & \colhead{DEC(2000)}&
\colhead{velocity (km s$^{-1}$) }& \colhead{ error(km s$^{-1}$) }& \colhead{notes}}
\startdata
   1 &   6 52   29.03 &  69 23    8.62 &  32510 &  50 & member \\
   2 &   6 52   33.43 &  69 16    3.14 &  36680 &  40 & \\
   3 &   6 52   33.44 &  69 15   26.07 &  30970 &  40 & member \\
   4 &   6 52   35.33 &  69 23   40.20 & 168400 & 190 & \\
   5 &   6 52   36.55 &  69 22    3.14 &  29630 &  90 & member \\
   6 &   6 52   37.63 &  69 19    3.54 &  31980 &  70 & member \\
   7 &   6 52   37.82 &  69 21   54.51 &  32970 & 210 & member \\
   8 &   6 52   39.01 &  69 17    5.22 &  32750 &  40 & member \\
   9 &   6 52   40.66 &  69 18   50.44 &  32330 &  30 & member \\
  10 &   6 52   44.31 &  69 18   22.97 & 100500 & 100 & \\
  11 &   6 52   45.03 &  69 19   21.42 &  31130 &  80 & member \\
  12 &   6 52   46.01 &  69 22    0.91 &  34760 & 140 & member \\
  13 &   6 52   48.14 &  69 19   37.27 &  32840 &  50 & member \\
  14 &   6 52   48.72 &  69 19   15.71 &  32890 &  60 & member \\
  15 &   6 52   49.30 &  69 22   41.37 &  99580 &  40 & \\
  16 &   6 52   51.56 &  69 21    2.88 &  32010 &  80 & member \\
  17 &   6 52   53.15 &  69 24   15.00 &  62220 &  30 & \\
  18 &   6 52   57.00 &  69 24    9.04 &  73360 &  40 & \\
  19 &   6 52   58.64 &  69 24   24.01 &  31330 &  50 & member \\
  20 &   6 52   59.16 &  69 16   45.22 &  32630 &  90 & member \\
  21 &   6 52   59.82 &  69 19   29.41 &  31150 & 130 & member \\
  22 &   6 52   59.98 &  69 16   57.06 &  32380 &  50 & member \\
  23 &   6 53    1.18 &  69 20   42.80 &  77020 &  90 & \\
  24 &   6 53    1.45 &  69 17   16.97 &  32070 &  40 & member \\
  25 &   6 53    2.27 &  69 20   24.73 &  32740 &  30 & member \\
  26 &   6 53    2.94 &  69 17   12.44 &  31040 &  80 & member \\
  27 &   6 53    5.67 &  69 23    5.02 & 111500 &  30 & \\
  28 &   6 53    6.17 &  69 16   25.75 &  26360 &  30 & \\
  29 &   6 53    7.69 &  69 15   46.61 &  33350 &  60 & member \\
  30 &   6 53    7.82 &  69 19   41.20 &  32190 &  50 & member \\
  31 &   6 53    8.93 &  69 20   42.03 &  32180 &  40 & member \\
  32 &   6 53   10.03 &  69 20    3.09 &  32830 &  80 & member \\
  33 &   6 53   10.07 &  69 19   33.67 &  31870 &  70 & member \\
  34 &   6 53   10.55 &  69 19    7.63 &  31980 & 100 & member \\
  35 &   6 53   13.39 &  69 20   28.74 &  33510 &  70 & member \\
  36 &   6 53   13.72 &  69 17   35.13 &  34270 &  50 & member \\
  37 &   6 53   14.07 &  69 19   49.79 &  33460 &  40 & member \\
  38 &   6 53   14.47 &  69 15   22.25 &  32800 &  70 & member \\
  39 &   6 53   15.05 &  69 20   14.21 &  31810 &  40 & member \\
  40 &   6 53   15.25 &  69 22    5.20 &  30390 & 140 & member \\
  41 &   6 53   16.05 &  69 17   55.67 &  32970 &  90 & member \\
  42 &   6 53   16.06 &  69 17   18.02 &  32910 &  70 & member \\
  43 &   6 53   16.38 &  69 20    5.31 &  32480 &  60 & member \\
  44 &   6 53   16.71 &  69 23   22.27 &  33840 &  40 & member \\
  45 &   6 53   16.77 &  69 17   46.72 &  32740 &  90 & member \\
  46 &   6 53   18.32 &  69 18    9.90 &  29350 &  60 & \\
  47 &   6 53   19.51 &  69 15   30.95 &  26360 &  60 & \\
  48 &   6 53   20.06 &  69 19   31.69 &  31320 &  60 & member \\
  49 &   6 53   20.87 &  69 16   11.63 &  33350 &  70 & member \\
  50 &   6 53   21.45 &  69 19   51.77 &  32750 &  60 &  WAT \\
  51 &   6 53   21.88 &  69 15   58.15 &  33820 &  60 & member \\
  52 &   6 53   22.29 &  69 21   33.12 &  30890 &  50 & member \\
  53 &   6 53   22.91 &  69 23   48.06 &  33190 &  30 & member \\
  54 &   6 53   23.11 &  69 23   43.25 &  28150 &  40 & \\
  55 &   6 53   23.82 &  69 15   57.32 &  32420 &  90 & member \\
  56 &   6 53   24.57 &  69 23   39.74 &  61950 &  70 & \\
  57 &   6 53   24.91 &  69 23   37.57 &  17020 &  40 & \\
  58 &   6 53   25.35 &  69 19   54.27 &  32970 & 100 & member \\
  59 &   6 53   26.12 &  69 18    8.72 &  32420 &  40 & member \\
  60 &   6 53   26.85 &  69 16   33.14 &  51640 &  70 & \\
  61 &   6 53   27.19 &  69 19   53.20 &  32230 &  70 & member \\
  62 &   6 53   27.56 &  69 18   41.62 &  32570 &  80 & member \\
  63 &   6 53   28.66 &  69 20    9.79 &  32960 &  90 & member \\
  64 &   6 53   31.52 &  69 23   29.88 &  32730 &  50 & member \\
  65 &   6 53   31.85 &  69 20   53.49 & 169300 &  60 & \\
  66 &   6 53   31.90 &  69 17   32.35 &  34560 &  60 & member \\
  67 &   6 53   34.69 &  69 19   18.70 &  33330 &  70 & member \\
  68 &   6 53   35.66 &  69 19   55.75 &  32840 &  70 & member \\
  69 &   6 53   37.11 &  69 16   32.42 &  30900 &  70 & member \\
  70 &   6 53   37.16 &  69 19   13.92 &  32510 &  90 & member \\
  71 &   6 53   37.46 &  69 19   22.30 &  32750 &  80 & member \\
  72 &   6 53   41.46 &  69 17    8.57 &  92040 &  80 & \\
  73 &   6 53   41.59 &  69 21   25.35 &  32710 &  40 & member \\
  74 &   6 53   42.33 &  69 17    1.18 & 175300 &  70 & \\
  75 &   6 53   43.96 &  69 18   13.96 &  31950 &  70 & member \\
  76 &   6 53   45.01 &  69 15   37.85 &  89980 & 110 & \\
  77 &   6 53   47.11 &  69 22   11.13 &  32300 &  90 & member \\
  78 &   6 53   47.18 &  69 21   16.64 &  34520 &  60 & member \\
  79 &   6 53   47.90 &  69 15   13.57 &  30450 &  80 & member \\
  80 &   6 53   48.49 &  69 24   23.35 &  89680 &  40 & \\
  81 &   6 53   49.66 &  69 20   26.84 &  32030 & 150 & member \\
  82 &   6 53   50.45 &  69 19    5.63 &  33000 &  60 & member \\
  83 &   6 53   50.86 &  69 18    1.30 &  33170 & 230 & member \\
  84 &   6 53   54.05 &  69 21   56.90 &  34390 & 220 & member \\
  85 &   6 53   55.40 &  69 16    5.42 &  32040 &  70 & member \\
  86 &   6 53   55.51 &  69 21    7.19 &  33350 &  50 & member \\
  87 &   6 53   56.38 &  69 23    3.12 &  38760 &  50 & \\
  88 &   6 54    0.48 &  69 20   33.00 &  33000 &  40 & member \\
  89 &   6 54    0.51 &  69 15   53.67 &  44450 &  60 & \\
  90 &   6 54    1.60 &  69 22   32.53 &  31690 & 130 & member \\
  91 &   6 54    2.83 &  69 16   43.49 &  33170 &  70 & member \\
  92 &   6 54    4.90 &  69 18   35.83 &  62870 &  80 & \\
  93 &   6 54    6.92 &  69 18   26.38 &  33440 &  60 & member \\
  94 &   6 54    7.44 &  69 22   24.48 &  31770 &  40 & member \\
  95 &   6 54    8.26 &  69 17   13.35 & 124700 &  70 & \\
  96 &   6 54    8.62 &  69 15   27.71 & 139400 &  80 & \\
  97 &   6 54    8.63 &  69 21   47.35 &  32500 &  50 & member \\
  98 &   6 54   10.63 &  69 16   21.57 &  28600 & 200 & \\
  99 &   6 54   10.63 &  69 19   39.93 &  32960 &  70 & member \\
 100 &   6 54   10.77 &  69 16   54.09 & 124600 & 130 & \\
 101 &   6 54   14.99 &  69 16   39.15 &  30910 &  70 & member \\
 102 &   6 54   15.58 &  69 19   46.85 &  34510 & 190 & member \\
\enddata
\end{deluxetable}

\begin{deluxetable}{c c c}
\tablewidth{0pt}
\tablecaption{\label{tab:tab2d} Substructure Tests for the CS Sample}
\tablehead{
  \colhead{substructure test}& \colhead{statistical significance} & \colhead{sensitivity}}
\startdata
AST & 0.63 & clumping\\
symmetry test ($\beta$) & 0.95& asymmetry\\
fourier elongation & 0.93 & elongation\\
Lee 2D & 0.07& bimodality\\
\enddata
\tablecomments{Column 3 is adapted from \citet{1996ApJS..104....1P}} 
\end{deluxetable}

\begin{deluxetable}{c c c}
\tablewidth{0pt}
\tablecaption{\label{tab:tab3d} Substructure Tests for the 3D Sample}
\tablehead{
  \colhead{substructure test}& \colhead{statistical significance}& \colhead{sensitivity}}
\startdata
Lee 3D & 0.82 & bimodality\\
$\alpha$ & 0.79 & centroid shift with velocity\\
$\epsilon$ & 0.11 & change in velocity dispersion and density with position\\
$\Delta$ & 0.13 & change of velocity dispersion and mean velocity with position\\
\enddata
\tablecomments{Column 3 is adapted from \citet{1996ApJS..104....1P}} 
\end{deluxetable}

\begin{deluxetable}{c c c c c c}
\tablewidth{0pt}
\tablecaption{\label{tab:tabemm} GMM groups}
\tablehead{
  \colhead{group label}& \colhead{number of}& \colhead{center RA}& \colhead{center DEC}& \colhead{location} & \colhead{scale}\\
  \colhead{}& \colhead{galaxies} & \colhead{(J2000)}&
  \colhead{(J2000)}& \colhead{$km s^{-1}$} & \colhead{$km s^{-1}$}}
\startdata
A & 56 & 6 53 20.6 & 69 19 26.4 & 32,661 & 947 \\
B & 20 & 6 53 24.5 & 69 19 16.3 & 32,118 & 1,102\\
\enddata
\end{deluxetable}

\begin{figure}
\includegraphics[angle=-90, scale=.7]{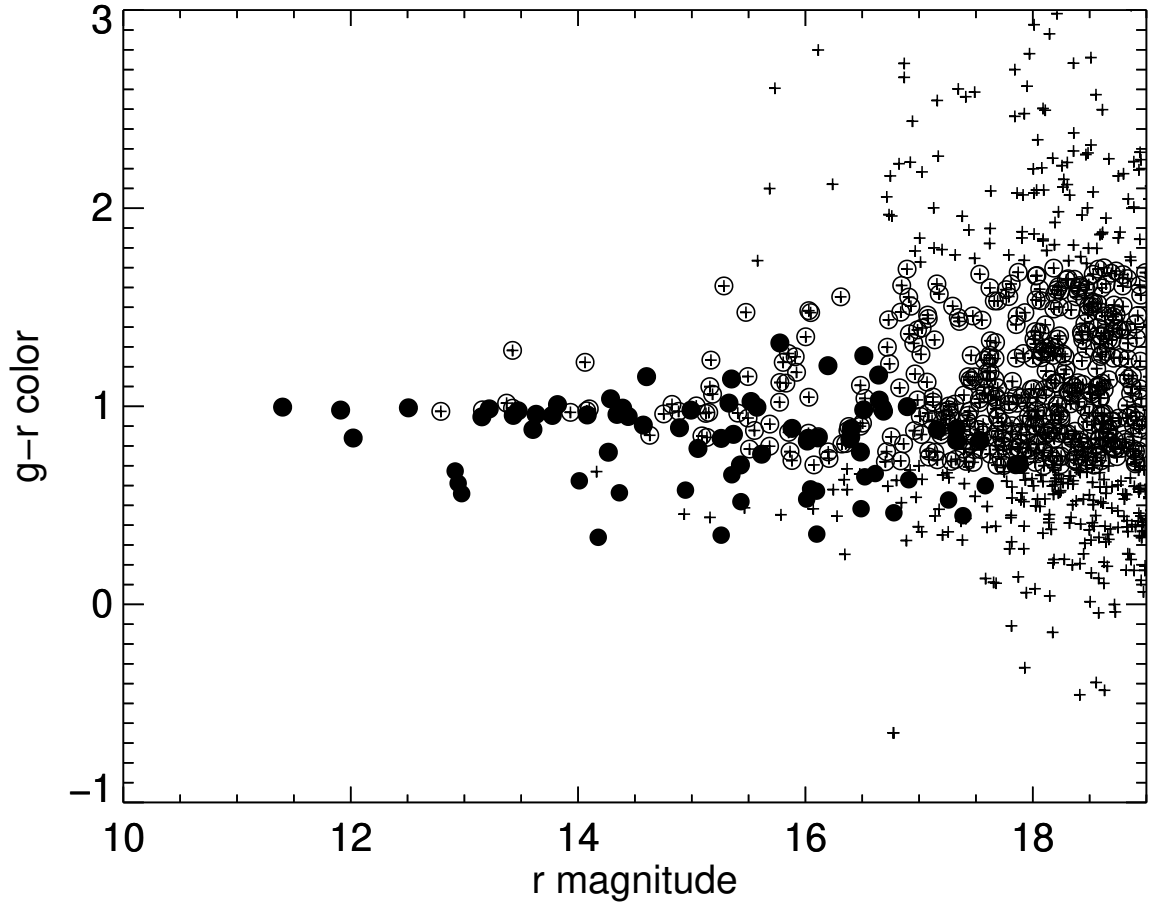}
\caption{The color-magnitude diagram derived from GMOS-N data for
  galaxies in the Abell 562 field. The crosses show all the likely
  galaxies in this field (CLASS-STAR $<$ 0.9). The circles show
  the cluster galaxy subsample that we derived (i.e., they had colors
  in the 0.3 $<$ g'-r' $<$ 1.7 range and were brighter than m$_{r'}$ of
  18.6) and that we could fit on the 4 masks. Of these, the solid circles show the spectroscopically confirmed cluster members.}
\label{fig:colmag}
\end{figure}

\begin{figure}
\plotone{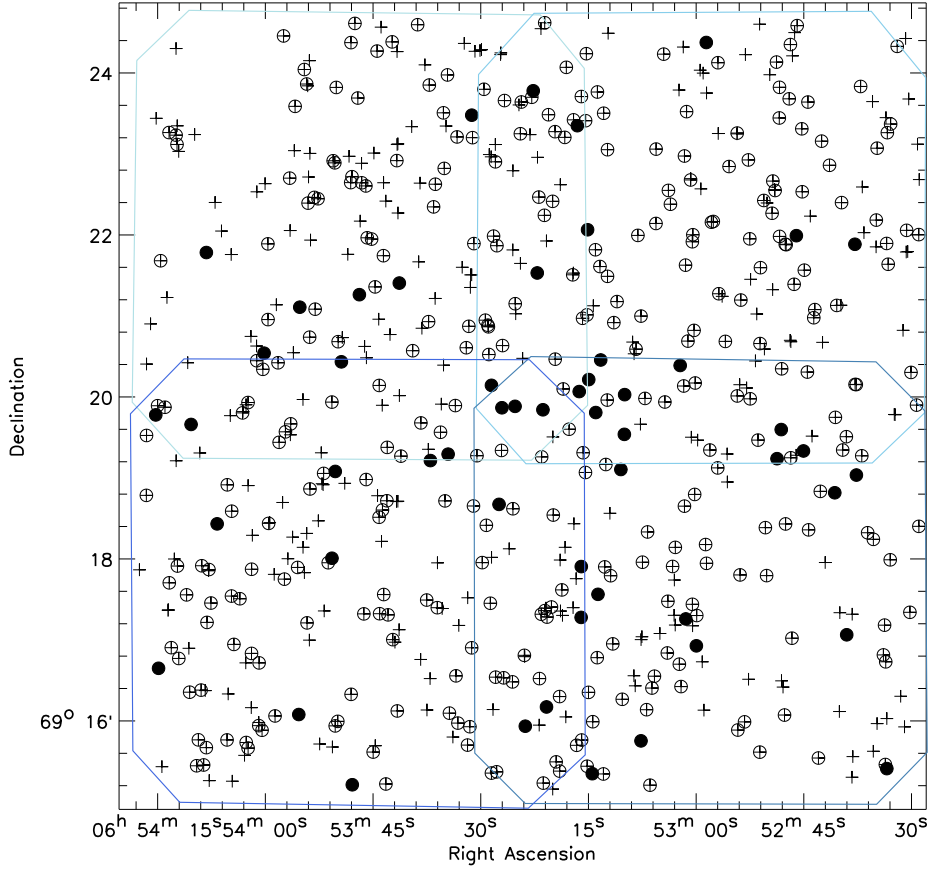}
\caption{The distribution of galaxies in the Abell 562 field derived from GMOS-N data. The crosses show all the likely
  galaxies in this field (CLASS-STAR $<$ 0.9). The circles show
  the cluster galaxy subsample that we derived (i.e., they had colors
  in the 0.3 $<$ g'-r' $<$ 1.7 range and were brighter than m$_{r'}$ of
  18.6) and that we could fit on the 4 masks. Of these, the solid circles show the spectroscopically confirmed cluster members. The blue octagons show the approximate edge of the field-of-view of each mask.}
\label{fig:galdist}
\end{figure}

\begin{figure}
\plottwo{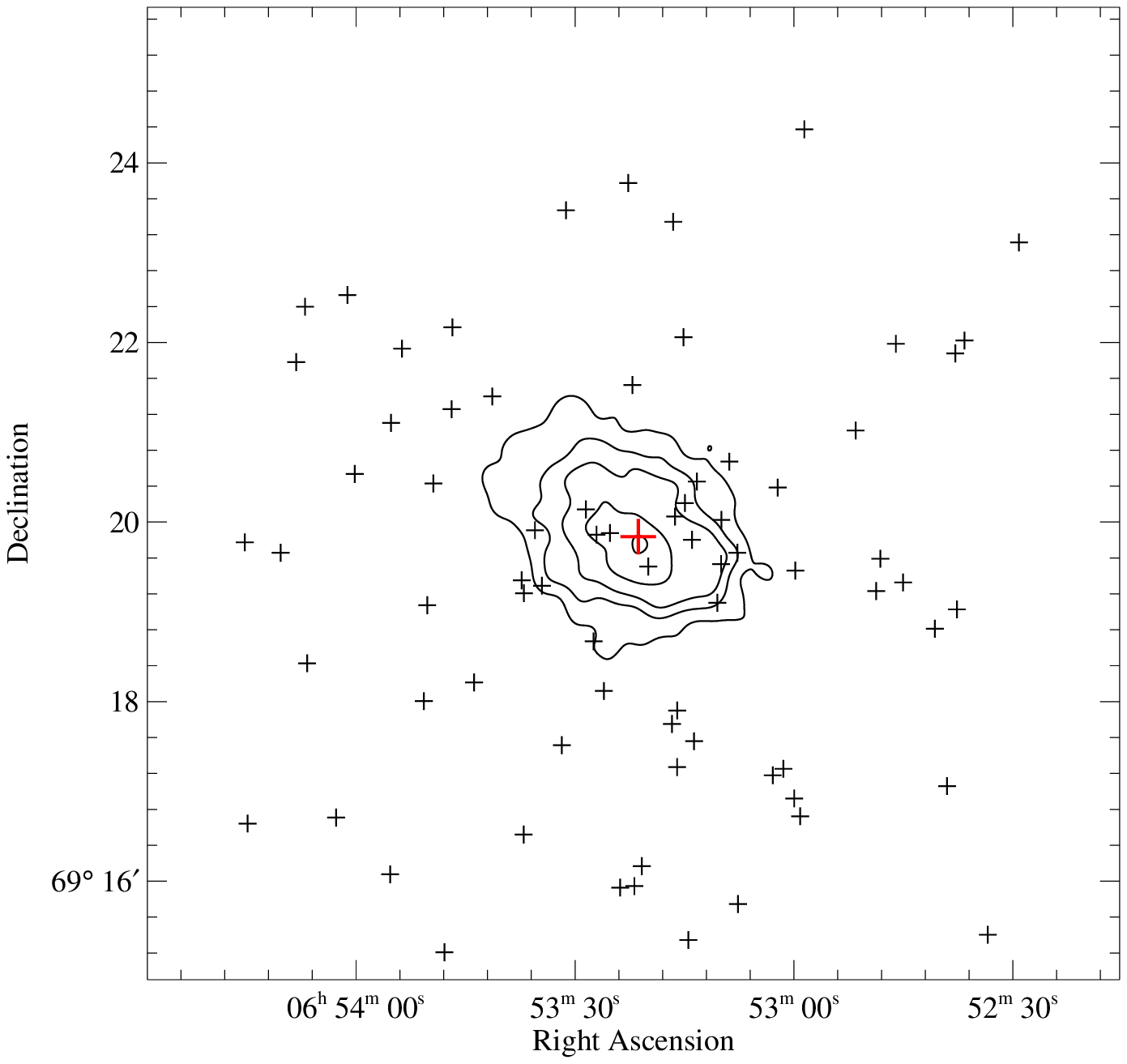}{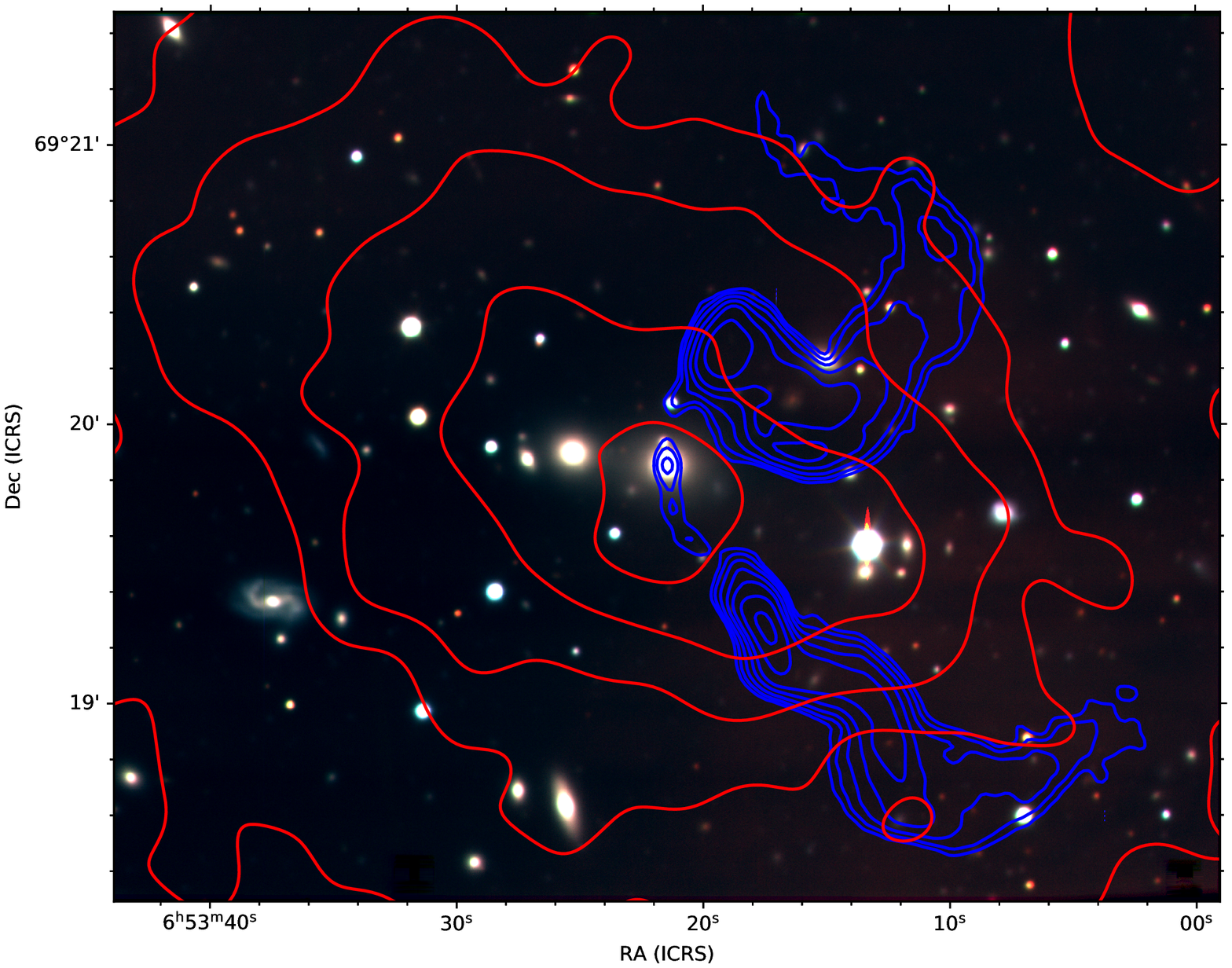}
\caption{Left. Contours of an adaptively smoothed Chandra image of the cluster in the 0.2-10 keV energy range.
The black crosses show the positions of the 76
  spectroscopically confirmed cluster members. Note that the X-ray emission is elongated in the NW-SE direction. The red cross marks the position of the WAT galaxy. Right. Deimos false color image (V, R, and I bands) of the central region of Abell 562 with overlaid X-ray (red) and 1.4 GHz VLA (in blue from E. M. Douglass private communication) contours. }
\label{fig:xrayimage}
\end{figure}

\begin{figure}
\includegraphics[angle=-90, scale=0.7]{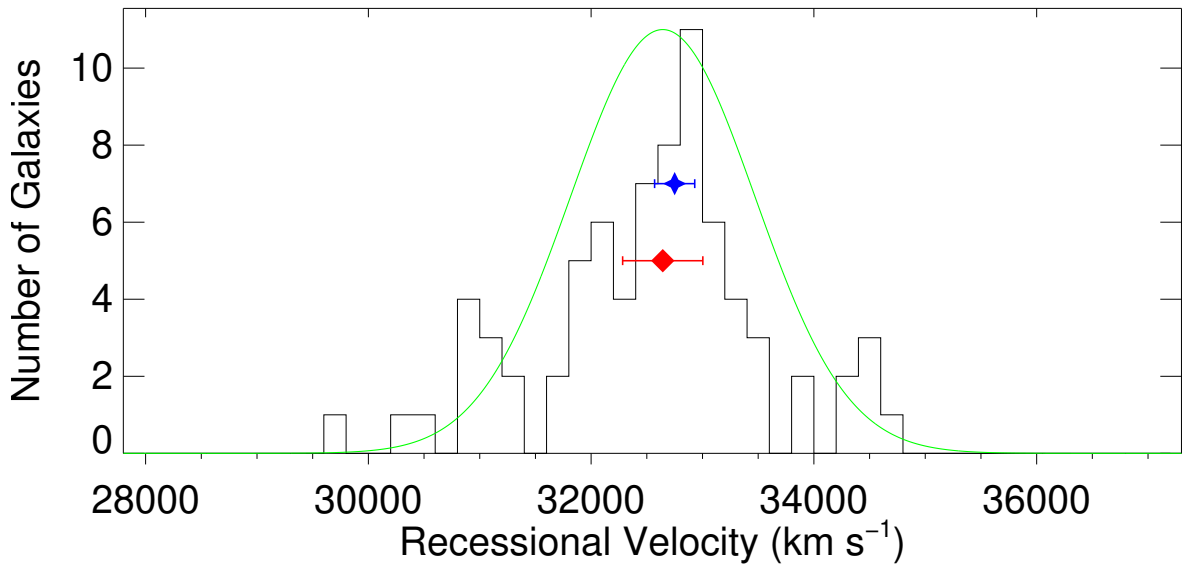}
\caption{Velocity histogram of the galaxies in Abell 562 The bin size is 250 km
  s$^{-1}$. The blue star shows the location of the WAT host and the red diamond shows the location of the robust bi-weight mean velocity. All the error bars show the 1$\sigma$ error. We also overlay (green) a Gaussian distribution derived from the measured bi-weight mean velocity and dispersion.}
\label{fig:velhist}
\end{figure}

\begin{figure}
\plotone{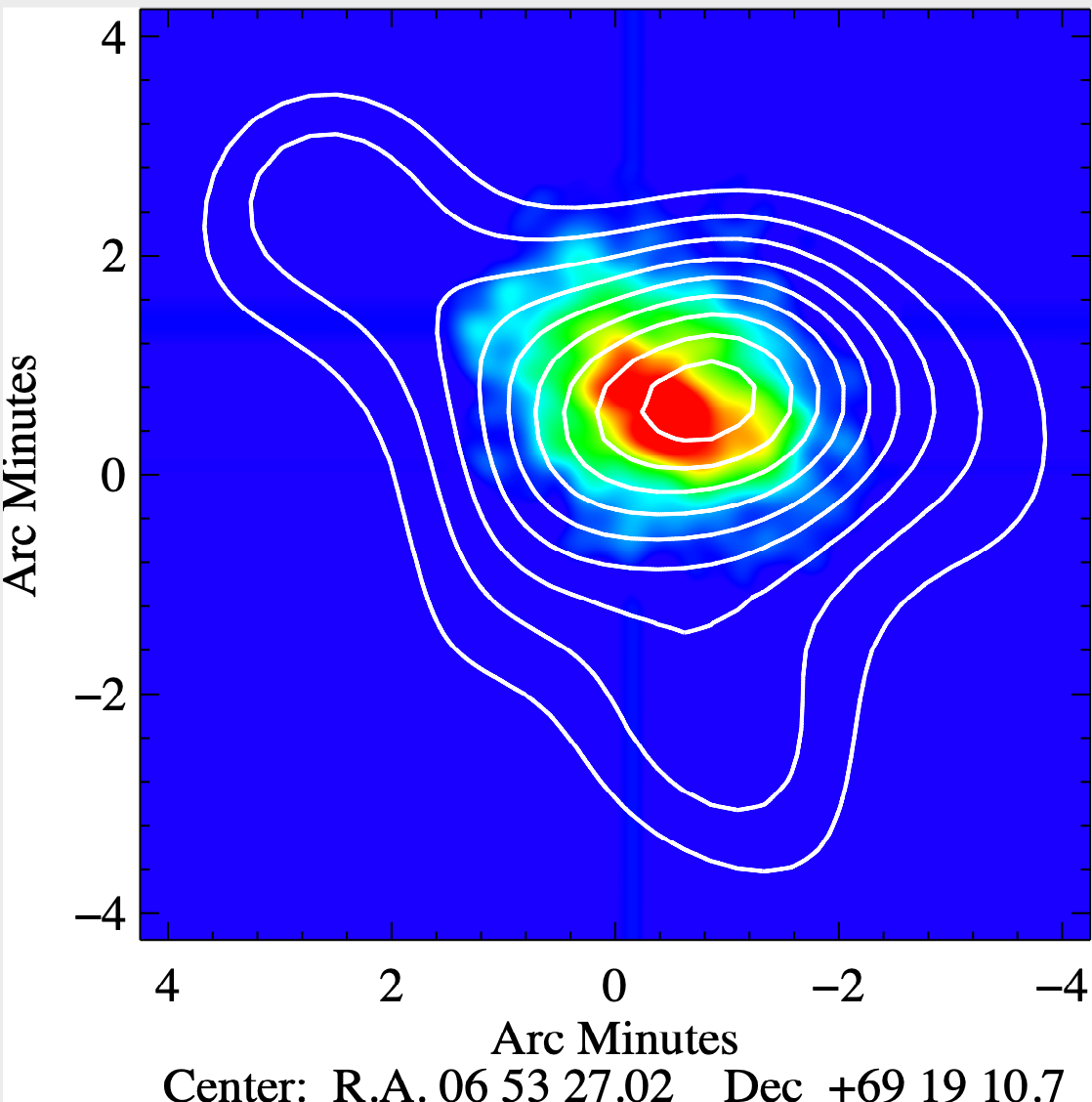}
\caption{An overlay of the spatial density of the 76 spectroscopically confirmed galaxy members  (white contours)
  over a color scale image of the adaptively smoothed 0.2-10 keV X-ray emission.  The contour values are 0.85, 1.1, 1.3, 1.5, 1.7, 1.9, 2.1, 2.4, and 2.6 galaxies arcmin$^{-2}$. Note that
  the X-ray peak is slightly offset from the peak in the galaxy
  density and that the X-ray emission is elongated.}
\label{fig:galdensityv}
\end{figure}

\begin{figure}
\plotone{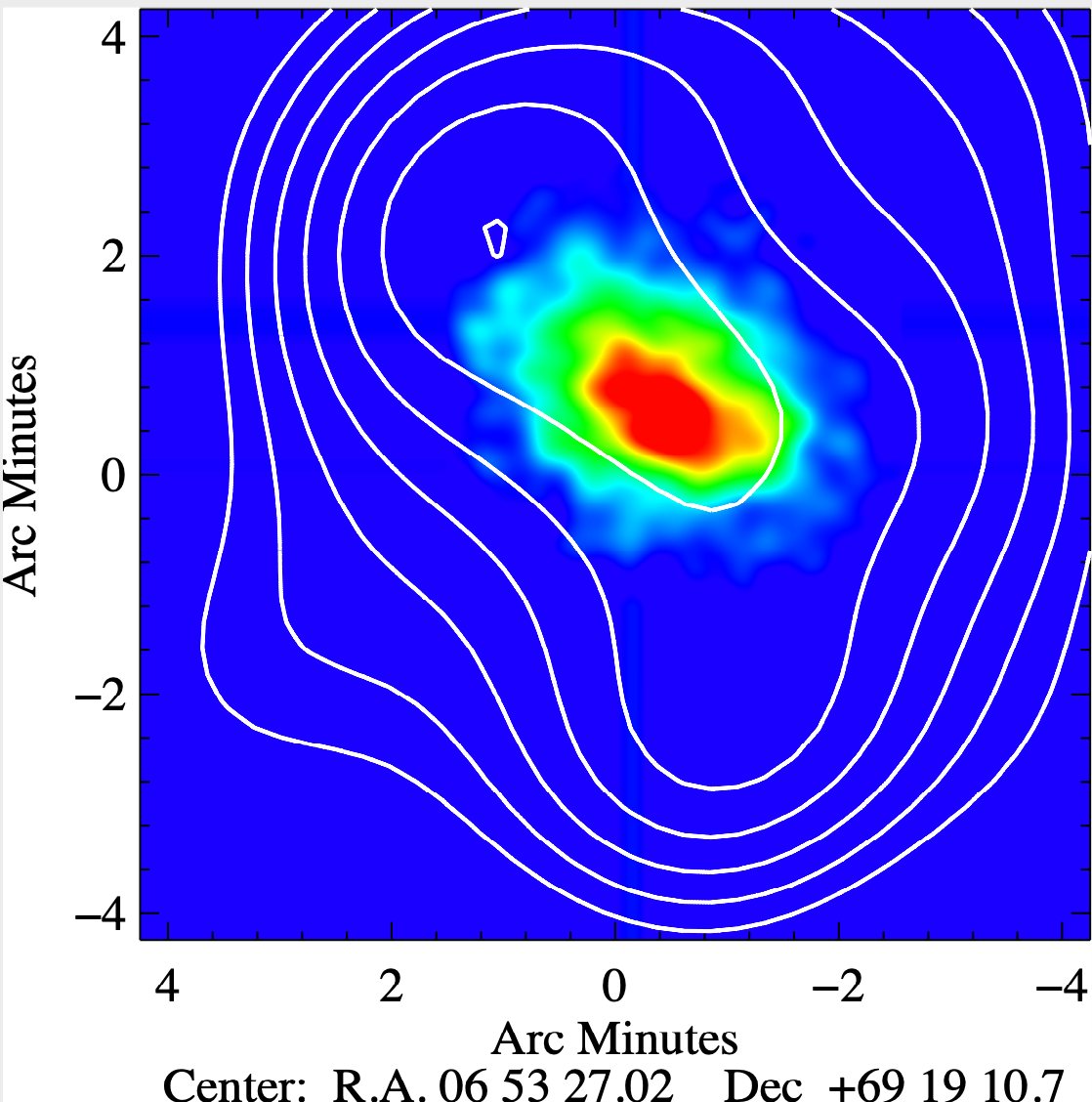}
\caption{An overlay of the spatial density of the 316 galaxies selected based on their color as likely cluster members  (white contours)
  over a color scale image of the adaptively smoothed 0.2-10 keV X-ray emission.  The contour values are 2.3, 2.6, 2.9, 3.2, 3.5 3.8, and 4.2 galaxies arcmin$^{-2}$. Note that
  the X-ray peak is aligned with the peak in the galaxy
  density and that the X-ray emission is elongated.}
\label{fig:galdensityc}
\end{figure}

\begin{figure}
\plotone{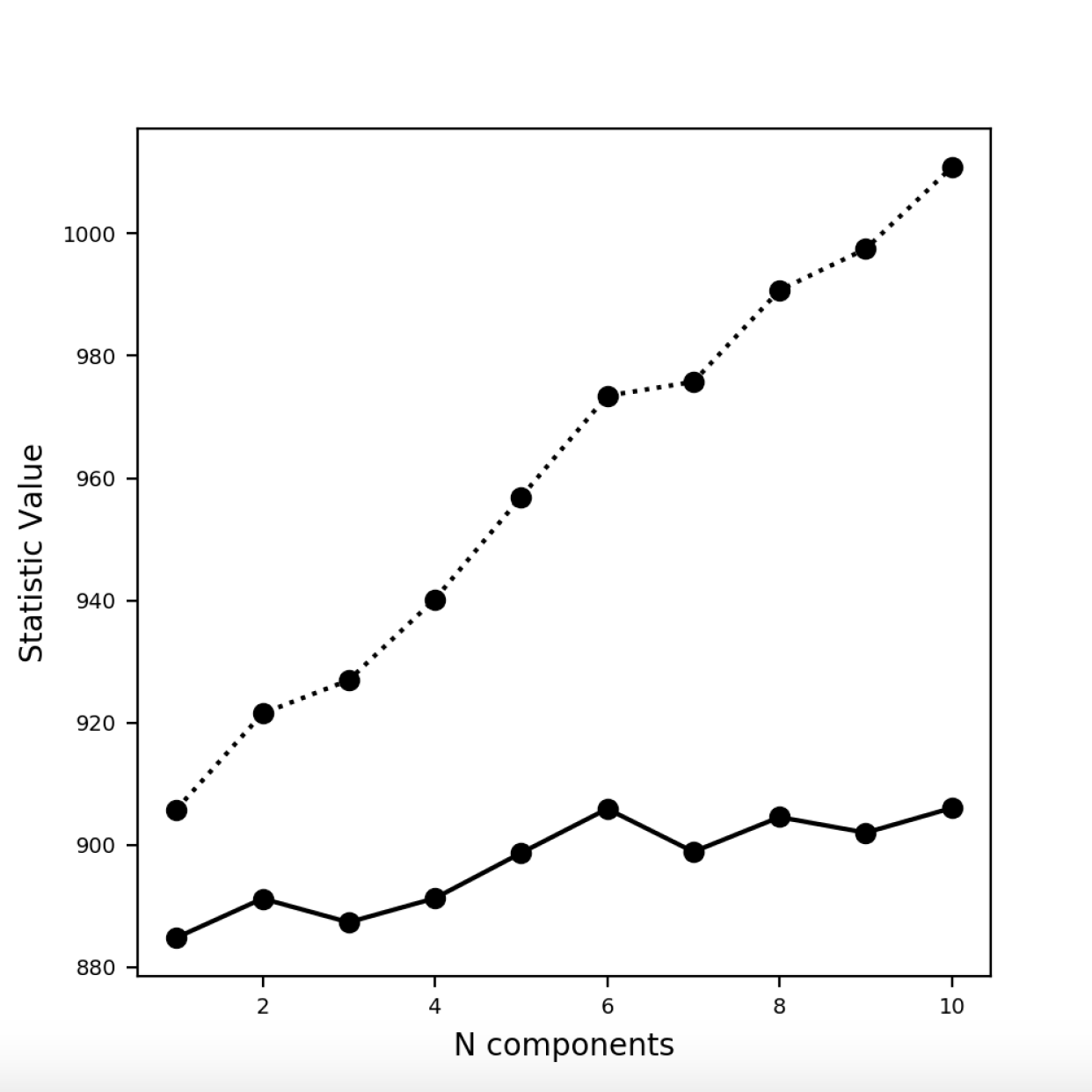}
\caption{Model selection criteria AIC (solid line) and BIC (dotted line) as a function of the number of components for the sample that combines positions and velocities (3D data set).}
\label{fig:ABstat}
\end{figure}

\begin{figure}
\plottwo{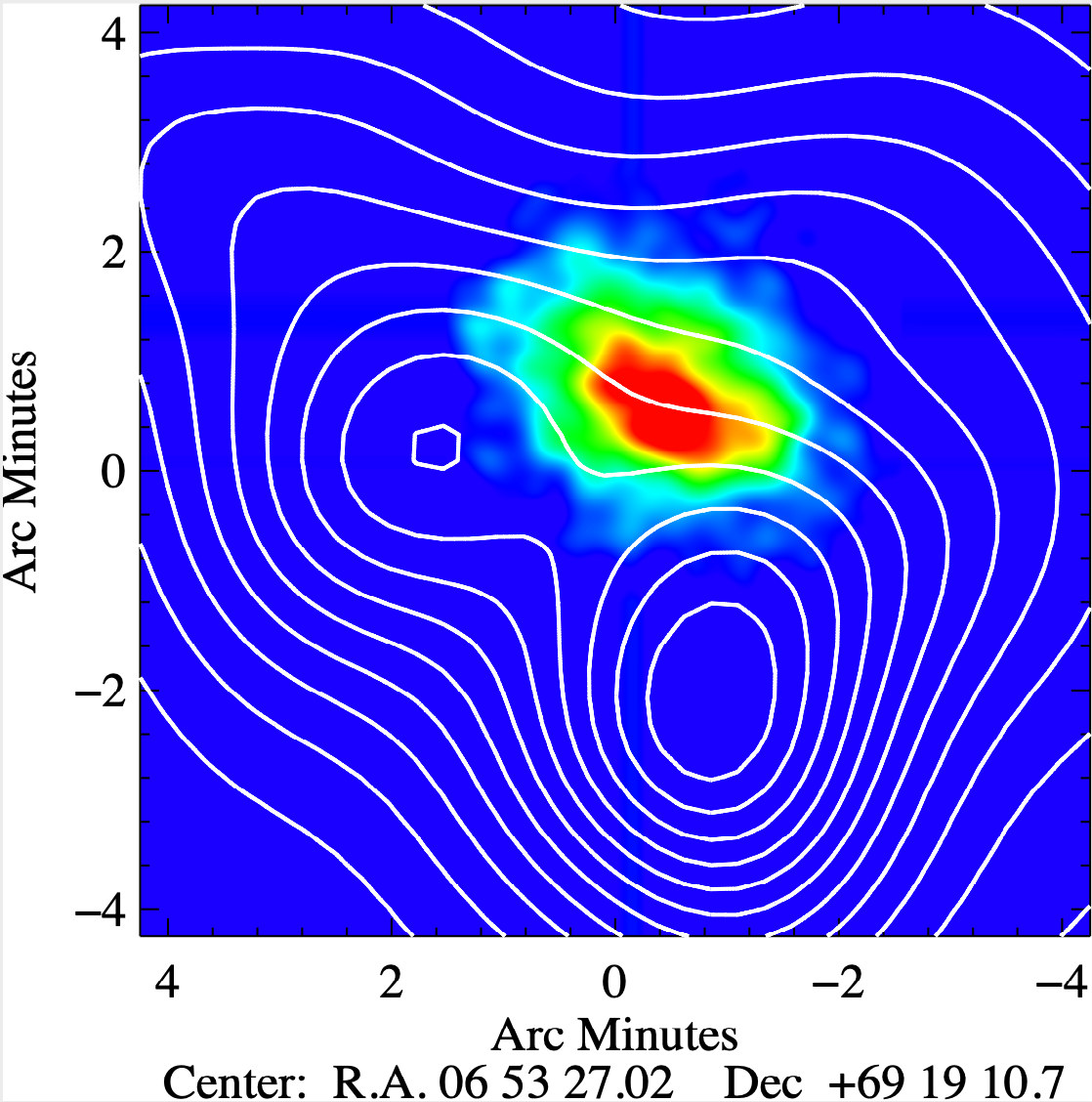}{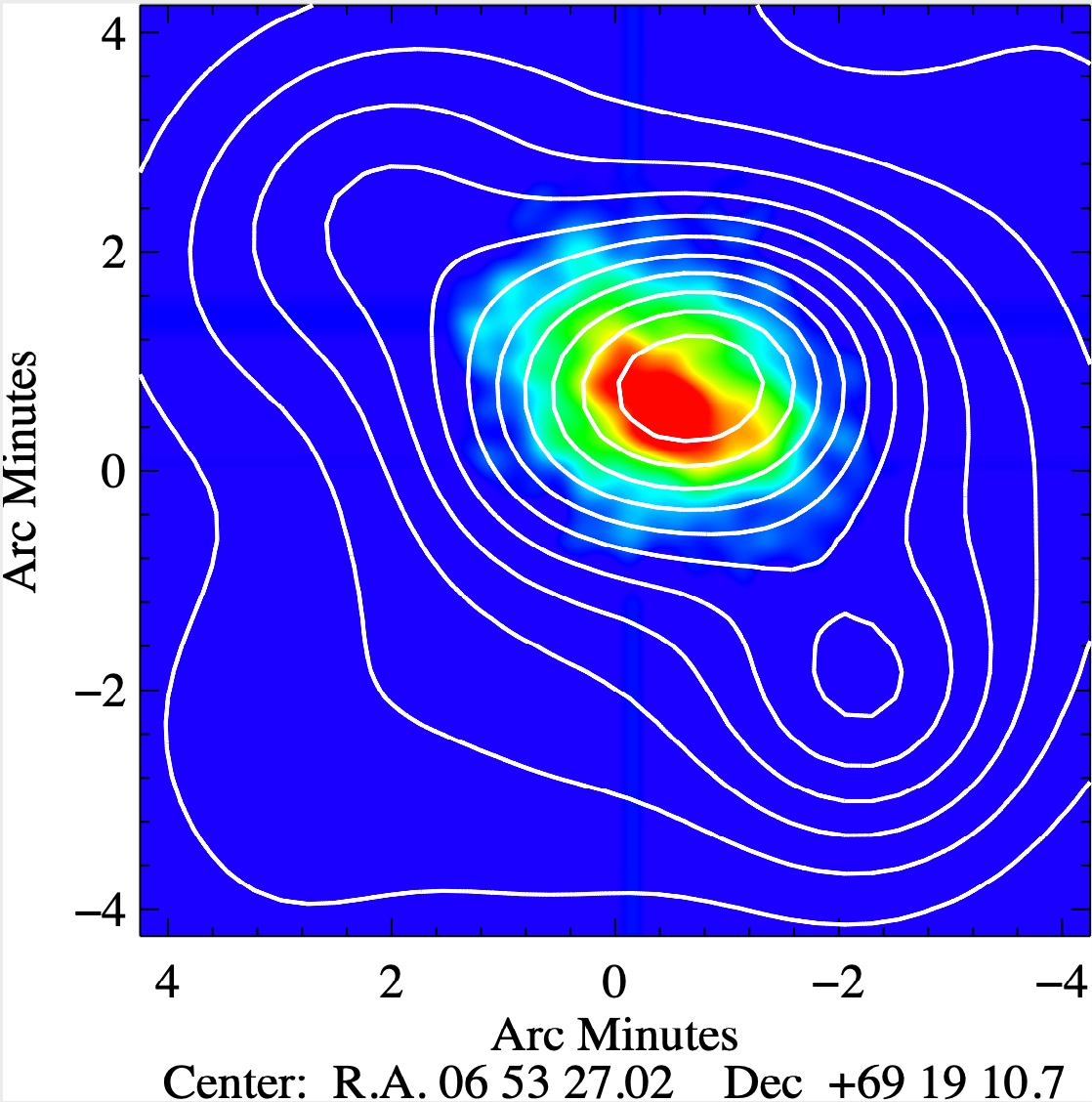}
\caption{Overlay of the galaxy densities as derived from the GMM models over a color scale image of the adaptively smoothed 0.2-10 keV X-ray emission. The right image shows the more compact galaxy group A and the contours are 0.17, 0.34, 0.51, 0.68, 0.85, 1.1, 1.4, 1.57, 1.64, 1.81 and 2.1 galaxies arcmin$^{-2}$. The left image shows the most spatially dispersed group B with contours 0.04, 0.09, 0.13, 0.17, 0.21, 0.26, 0.3, 0.34, 0.39, 0.43}
\label{fig:2demm}
\end{figure}

\begin{figure}
\plotone{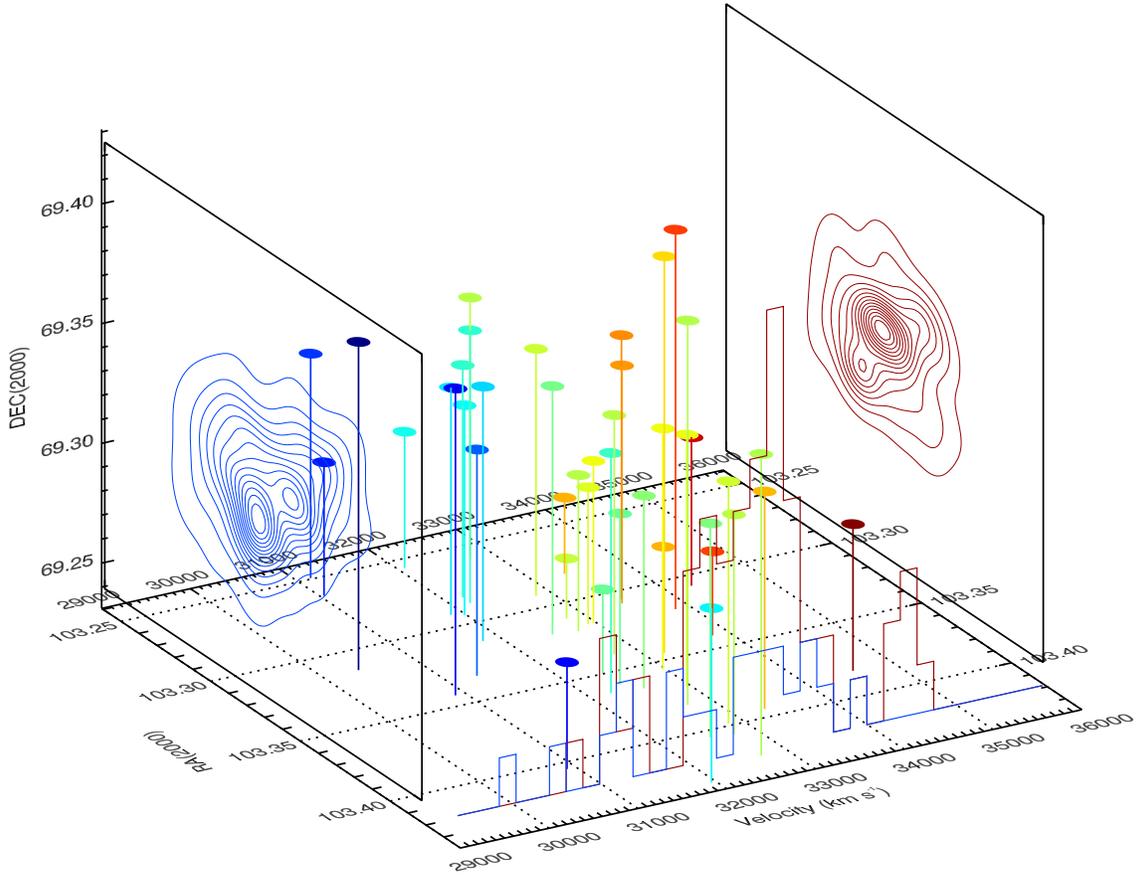}
\caption{3D scattered plot of the sky position and redshift of the cluster  members. The galaxies are color coded according to their velocities.  In addition, we have projected the velocity histograms of the two GMM groups. Group A in red and group B in blue.Finally, we also show the projected spatial density for the members of each GMM group. The contours of group A is in red and are projected to the right whereas the contours for group B are projected to the left.}
\label{fig:3dall}
\end{figure}

\begin{figure}
\plotone{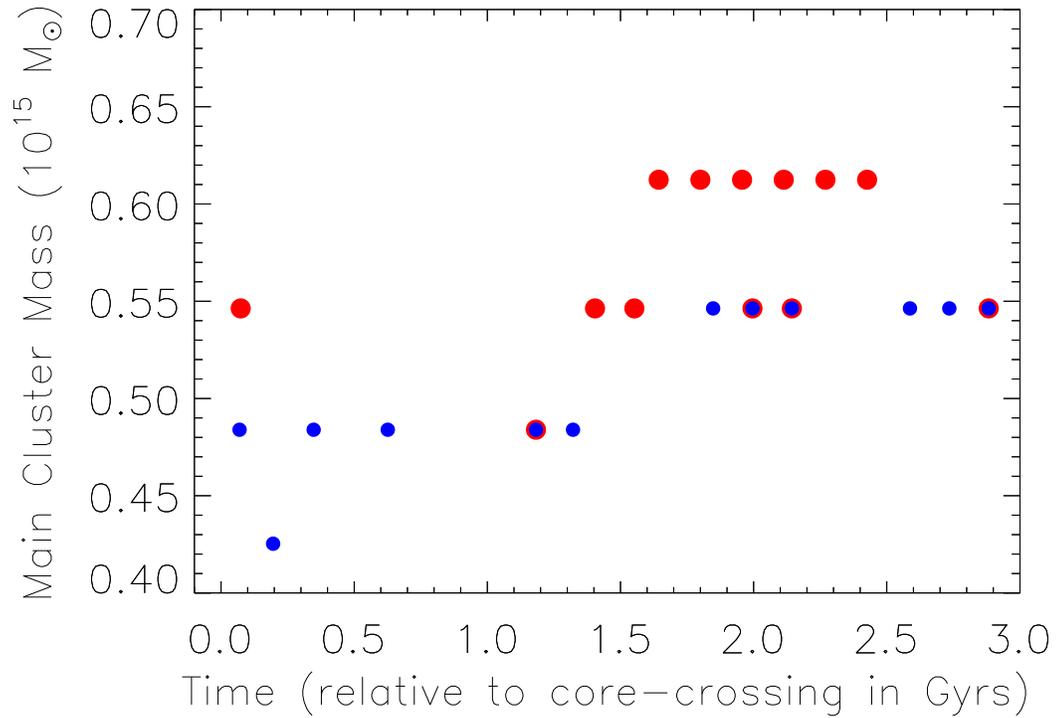}
\caption{Allowed merger models plotted as a function of main cluster mass and time since the epoch of closest approach, based on comparing the line-of-sight velocity distributions from the observed galaxies near the center with the modeled distributions. Those models with a K-S probability greater than 70\% are shown; many other possible models were rejected. In this case we only analyzed non head on mergers with 1 to 4 mass ratios. The red dots show mergers with impact parameter of 125 kpc and the blue dots show mergers with impact parameter of 500 kpc. Moreover, we only include models with viewing angle of 15 $\deg$ (where 0$\deg$ corresponds to viewing along the merger axis).}
\label{fig:accsim}
\end{figure}

\begin{figure}
\plotone{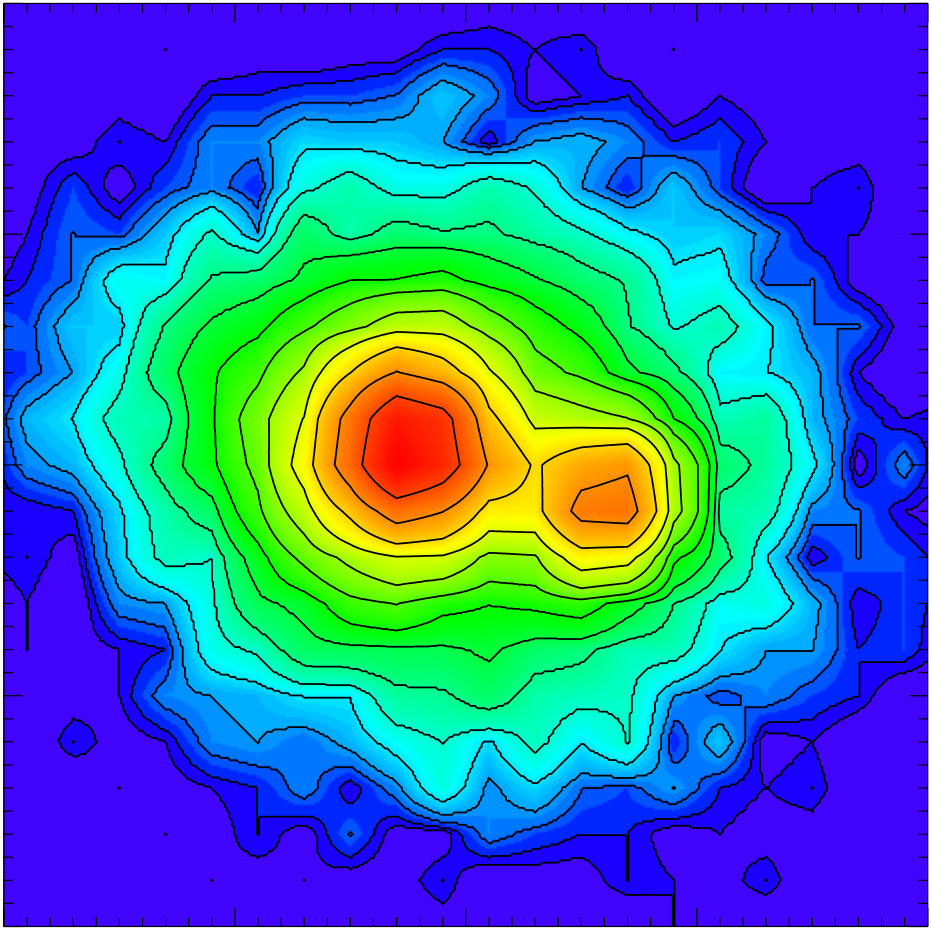}
\caption{Projected surface density plot for one of the simulated models (impact parameter of 500 kpc) depicted in Figure \ref{fig:accsim}. In this case the merger epoch is $\sim$ 0.5 Gyrs after the time of closest approach. The merger started from the left and the initial direction was parallel to the x axis. The projected distance between the two main clumps is $\sim$ 750 kpc which is comparable to the observed distance between the main clumps in Abel 562 of $\sim$ 600 kpc. The depicted region is 4.7 Mpc $\times$4.7 Mpc.}
\label{fig:contsim}
\end{figure}

\begin{figure}
\plotone{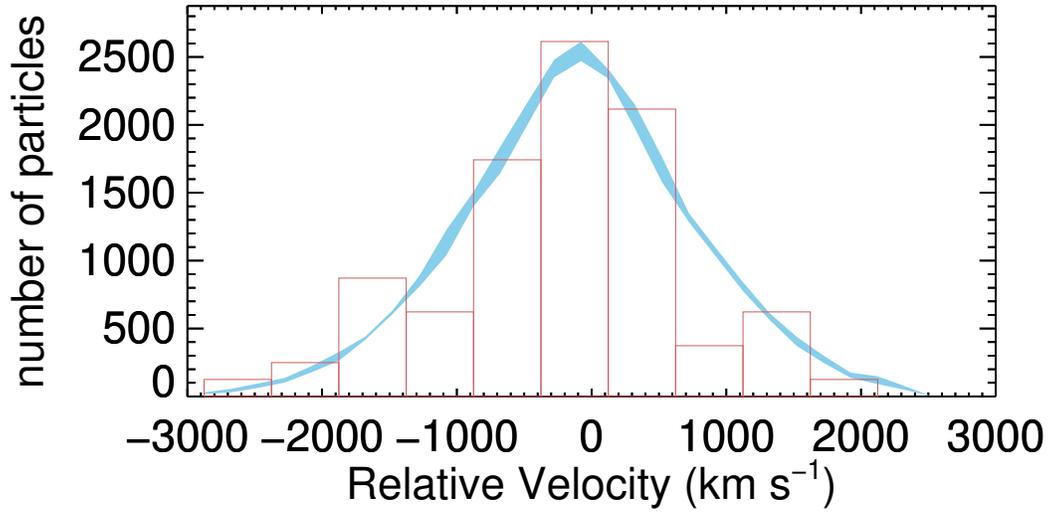}
\caption{Comparison of the observed and simulated velocity histograms. This figure shows the maximum and minimum envelope of the simulated histograms (blue) from  models with a Kolmorov-Smirnov probability greater than 0.7 (see figure \ref{fig:accsim}. The red plot shows the histogram of the observed velocities normalized to the maximum number of particles in the simulated models.}
\label{fig:plot_histsim}
\end{figure}

\begin{figure}
\plotone{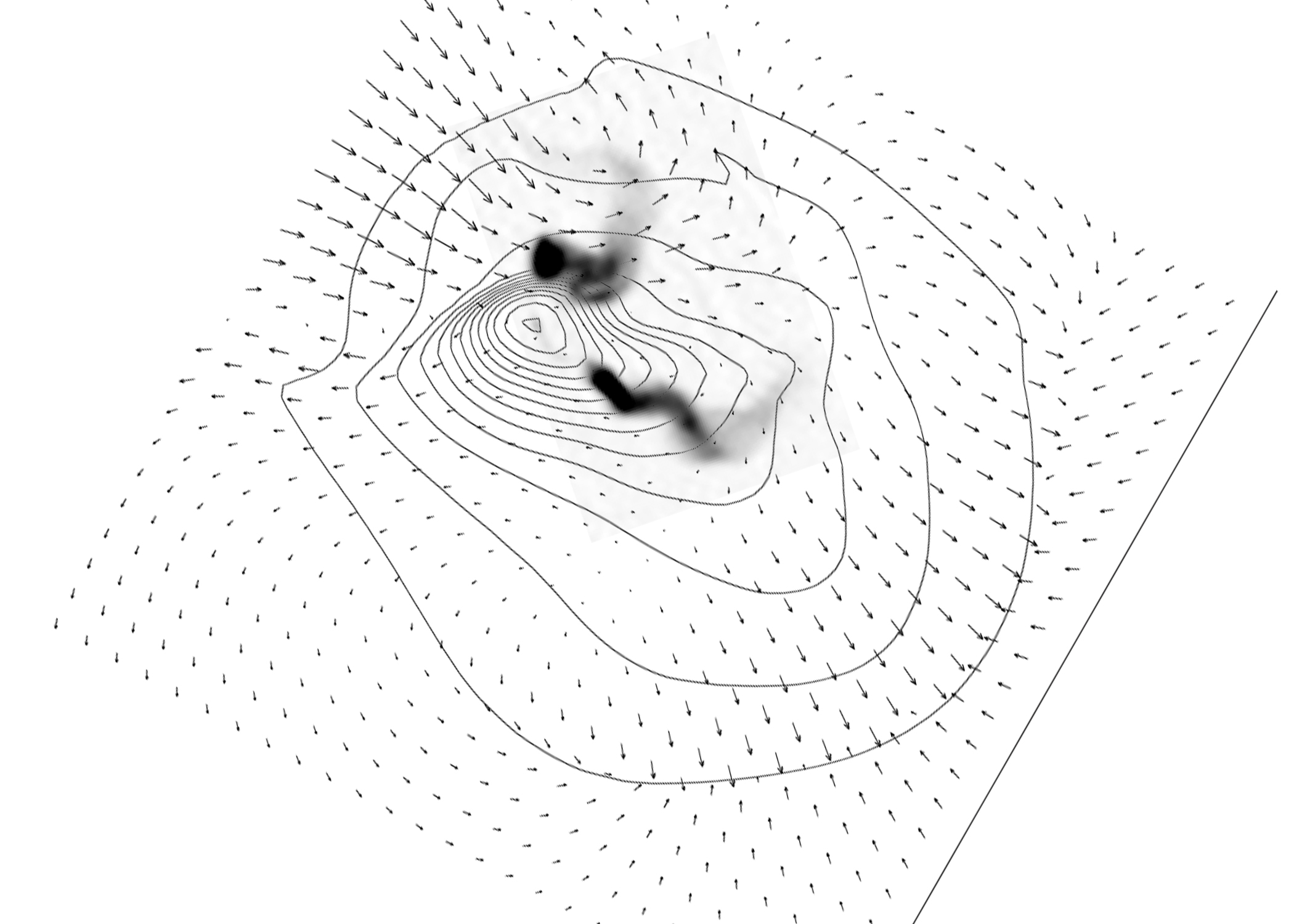}
\caption{\textbf{Overlay of a 1.4 GHz VLA radio map of the A562 WAT (from E. M. Douglass private communication)} over the velocity field predicted by the Abell 754 simulation (adapted from \cite{1998ApJ...493...62R}). The contours show the synthetic X-ray emission as predicted for a non head-on merger seen some 0.25Gyrs after the epoch of closest approach. Note that the arrows represent the gas velocity and the maximum value is 1850 $km s^{-1}$ in a $\sim 2Mpc \times 2Mpc$. We can see that the southern jet is oriented within a region free from large eddies whereas the northern jet encounters a fast rotating flow of gas and other eddies that could be responsible for the observed bends.}
\label{fig:plot_merged}
\end{figure}

\end{document}